\documentclass[aps,prb,twocolumn,superscriptaddress,reftest]{revtex4}

\usepackage{graphicx}
\usepackage{graphics}
\usepackage{ifthen,calc}
\usepackage{amsfonts}
\usepackage{amssymb}
\usepackage{amsmath}
\usepackage{subfigure}
\usepackage{bbm}

\def\rp#1{\left(#1\right)}
\def\kp#1{\left\{#1\right\}}
\def\sqp#1{\left[#1\right]}
\def\bp#1{\left|#1\right|}

\def\ket#1{|{#1}\rangle}

\def\ketbra#1#2{|{#1}\rangle\langle{#2}|}
\def\pro#1{\langle{#1}\rangle}

\def\matrix22#1#2#3#4{\left(\begin{array}{cc}{#1} & {#2}\\{#3} & {#4}\end{array}\right)}

\def\Ctilde{\tilde{\mathcal{C}}}
\def\C{\mathcal{C}}

\begin{document}


\title{Electron spin decoherence of single Nitrogen-Vacancy defects in diamond}


\author{J. R. Maze}
\affiliation{Department of Physics, Harvard University, 17 Oxford
St., Cambridge, MA 02138} %
\author{J. M. Taylor}
\affiliation{Department of Physics, Massachusetts Institute of
Technology, 77  Massachusetts Avenue, Cambridge, MA 02138}
\author{M. D. Lukin}
\affiliation{Department of Physics, Harvard University, 17 Oxford
St., Cambridge, MA 02138}


\date{\today}

\begin{abstract}
  We present a theoretical analysis of the electron spin decoherence in single Nitrogen-Vacancy defects in ultra-pure diamond. The electron spin decoherence is due to the interactions with Carbon-13 nuclear spins in the diamond lattice. Our approach takes advantage of the low concentration (1.1\%) of Carbon-13 and their random distribution in the diamond lattice by an algorithmic aggregation of spins into small, strongly interacting groups.  By making use of this \emph{disjoint cluster}  approach, we demonstrate a possibility of non-trival dynamics of the electron spin that can not be described by a single time constant. This dependance is caused by a strong coupling between the electron and few nuclei and results, in particular, in a substantial echo signal even at microsecond time scales. Our results are in good agreement with recent experimental observations.
\end{abstract}


\maketitle

\section{Introduction}

Isolated spins in solid-state systems are currently being explored as candidates for good quantum bits, with applications to quantum computation~\cite{kane1998,loss1998,wrachtrup2004}, quantum communication\cite{childress2006prl} and magnetic sensing\cite{taylor2008, maze2008nature, balasubramanian2008nature}. The Nitrogen-vacancy (NV) center in diamond is one such isolated spin system. It can be prepared and detected using optical fields, and microwave radiation can be used to rotate the spin~\cite{manson1990,jelezko2004}. Recent experiments have conclusively demonstrated that in ultra pure-diamond the electron spin coherence lifetime is limited by its hyperfine interactions with the natural 1.1\% abundance Carbon-13 in the diamond crystal~\cite{vanoort1990, childress2006}. Thus, developing a detailed understanding of the decoherence properties of such an isolated spin in a dilute spin bath is a challenging problem of immediate practical interest. This combined system of electron spin coupled to many nuclear spins has a rich and complex dynamics associated with many-body effects.

The decay of electronic spin coherence due to interactions with surrounding nuclei has been a subject of a number of theoretical studies~\cite{shenvi2005, coish2004}. Various mean-field and many-body approaches have been used to address this problem~\cite{desousa2003prb67,desousa2003,witzel2005, coish2005, yao2006, saikin2007}. In this paper, we investigate a variation of the cluster expansion, developed in Ref.~\onlinecite{witzel2005}.  Our approach takes advantage of the natural grouping statistics for randomly located, dilute impurities, which leads to the formation of small, disjoint clusters of spins which interact strongly within themselves and with the central spin, but not with other such clusters.  This suggests a natural hierarchy of interaction scales of the system, and allows for a well-defined approximation that can be seen as an extension of ideas developed in the study of tensor networks~\cite{shi2006}.  We develop an algorithm for finding clusters given a set of locations and interactions, and find that for dilute systems convergence as a function of the cluster size (number of spins in a given cluster) is very rapid.  We then apply this technique to the particular problem of the decay of spin-echo for a single NV center, and find good qualitative and quantitative agreement with experiments. In particular, we demonstrate a possibility of non-trival dynamics of the electron spin that can not be described by a single time constant. This dependance is caused by a strong coupling between the electron and few nuclei and results in a substantial spin-echo signal even at microseconds time scale.

\section{Methods}\label{sec:ham}

In this section, we introduce the properties of the electron spins in a NV center and the nuclear spins in its environment. Then, we address the many body problem involved in the evaluation of spin-echo signals.

\subsection{Spin hamiltonian}

The negatively charged NV center ([N-V]$^-$) has trigonal $C_{3v}$ symmetry and $^3A_2$ ground state\cite{goss1996} with total electronic spin $S=1$\cite{he1993}. Spin-spin interaction leads to a
zero-field splitting, $\Delta = 2.87$ GHz, between the $m_s=0$ and $m_s=\pm1$ manifolds, where the quantization axis is along the NV-axis. This spin triplet interacts via hyperfine interaction with a spin bath composed of the adjacent Nitrogen-14 and the naturally occuring 1.1\% Carbon-13 which is randomly distributed in the diamond lattice.

In the presence of an external magnetic field, the dynamics is governed by the following hamiltonian,
\begin{eqnarray}
\label{eq:ham}%
H & = & \Delta S_z^2 - \gamma_eB_zS_z -
\sum_n\gamma_N\mathbf{B}\cdot \mathbf{g}_n(|S_z|) \cdot \mathbf{I}_n \nonumber\\
& &
+\sum_{n}S_z\mathbf{A}_n\cdot\mathbf{I}_n + \sum_n \delta \mathbf{A}_n(|S_z|) \cdot \mathbf{I}_n \nonumber\\
& &+ \sum_{n>m}\mathbf{I}_n\cdot\mathbf{C}_{nm}(|S_z|)\cdot\mathbf{I}_m.
\end{eqnarray}
The relatively large zero-field splitting $\Delta$ (first term in Eq. (\ref{eq:ham})) does not allow the electron spin to flip and thus we can make the so called secular approximation, removing all terms which allow direct electronic spin flips. Non-secular terms have been included up to second order in perturbation theory, leading to the $|S_z|$ dependence of other terms in the Hamiltonian. The second and third terms are, respectively, the Zeeman interactions for the electron and the nuclei, the fourth term is the hyperfine interaction between the electron and each nucleus, the fifth term is an effective crystal-field splitting felt by the nuclear spins, and the last term is the dipolar interaction among nuclei. The specific terms for this hamiltonian are discussed in the Appendix.

For the NV center case, the nuclear $g$-tensor, $\mathbf{g}_n$, can be anisotropic and vary dramatically from nucleus to nucleus\cite{childress2006}. This leads to a non-trivial dynamics between the electron and an individual nucleus (electron-nuclear dynamics), and motivates a new approach for the case of a dilute bath of spins described below. In addition, the interaction between nuclei is enhanced by the presence of the electron of the NV center. The resulting effective interaction strength can exceed several times the bare dipolar interaction between nuclei\cite{dutt2007}.

\subsection{Disjoint cluster method}\label{sec:mbi}

The large zero-field splitting $\Delta$ sets the quantization axis (called NV-axis) and allows us to neglect electron spin flips due  to interactions with nuclei. Therefore, we can reduce the Hilbert space of the system by projecting hamiltonian (\ref{eq:sham}) onto each of the electron spin states. We can write the projected hamiltonian, $P_{m_s}HP_{m_s}$ (where $P_{m_s} = \ketbra{m_s}{m_s}$), as
\begin{equation}
\label{eq:hms}%
H_{m_s} = \sum_n\mathbf{\Omega}_n^{({m_s})}\cdot\mathbf{I}_n +
\sum_{nm}\mathbf{I}_n\cdot\mathbf{C}_{nm}^{({m_s})}\cdot\mathbf{I}_m+\Delta|m_s|-\gamma_eB_zm_s,
\end{equation}%
where $m_s$ denotes the electron spin state, $\mathbf{\Omega}_n$ is the effective Larmor vector for nucleus $n$ and $\mathbf{C}_{nm}$ is the effective coupling between nuclei $n$ and $m$. In equation (\ref{eq:hms}), we include the zero-field splitting and the Zeeman interaction. These terms provide just static fields whose effect is canceled by spin echo. In this way, we can write the evolution of the bath as $U_{m_s}(\tau) = T\kp{\exp\rp{\int_{0}^{\tau} H_{m_s}\rp{t'}dt'}}$. An exact expression for $U_{m_s}$ can be found by ignoring the intra-bath interactions ($\mathbf{C}_{nm}=0$). However, for an interacting bath (with arbitrary $\mathbf{C}_{nm}$), solving $U_{m_s}$ for a large number of nuclei $N$ is a formidable task since it requires describing dynamics within a $2^N$ dimensional Hilbert space.  Therefore, some degree of approximation is needed.

\begin{figure}[t]
\includegraphics[width=0.7\columnwidth]{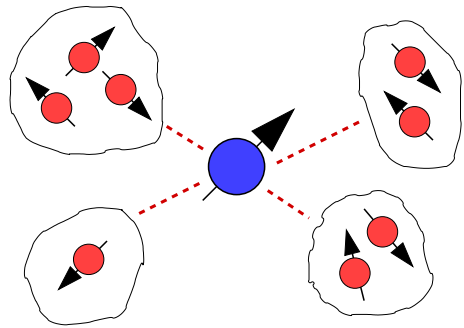}
\caption{Illustration of the method. Spins that strongly interact can be grouped together and treated as isolated systems. Interactions that joint different groups can be incorporated as a perturbation.}\label{figcluster}
\end{figure}

The spin bath considered here is composed of randomly distributed spins, and not all pair interactions among nuclei are equally important.  Specifically, interactions decay with a characteristic law $1/R_{mn}^3$, where $R_{nm}$ is the distance between nuclei $n$ and $m$. As a result, we can break the big problem into smaller ones by grouping those nuclei that strongly interact with each other into disjoint sets. Our procedure is illustrated in figure \ref{figcluster}. We denote the $k$-th group of nuclei as $\C_g^k$, where the subindex $g$ indicates that each group has no more than $g$ nuclei. Interactions inside each group (intra-group interactions) are expected to be much larger than interactions among groups (inter-group interactions). Our approximation method will rely upon neglecting the latter.

Formally, we start by separating intra-group interactions and inter-group interactions.  We define the operator $H_B = \sum_k H(\C_g^k)$ which contains all electron-nuclear interactions (first term
in Eq. (\ref{eq:hms})) plus all interactions between bath spins within the same group $\C_g^k$.  Similarly, operator $H_A = H(\Ctilde_g) (=H-H_B)$ contains all interactions between bath spins
in different groups. As a first approximation, we can neglect the inter-group interactions but keep the intra-group interactions. The approximation can be understood by means of the Trotter expansion\cite{trotter1959}
\begin{equation}
\label{eq:trotter}%
\exp\rp{H_At+H_Bt} = \lim_{n\rightarrow\infty}
\rp{\exp\rp{H_At/n}\exp\rp{H_Bt/n}}^n.\nonumber
\end{equation}
Since $H_B$ contains groups of terms that are disconnected from each other, $[H(\C_g^k),H(\C_g^{k^{\prime}})]=0$ and we can write the evolution operator as
\begin{equation}
\label{eq:utrotter}%
{U_g\rp{\tau} = \lim_{n\rightarrow\infty}
\rp{U\rp{\Ctilde_g,\frac{\tau}{n}}\prod_kU\rp{\C_g^k,\frac{\tau}{n}}}^n,}
\end{equation}
where $U(\Ctilde_g,\frac{\tau}{n})$ is the evolution operator due to hamiltonian $H(\Ctilde_g)$ and so on.

To the zeroth order we neglect all terms in $H_A$ since $H(\Ctilde_g)$ contains interactions among nuclei that interact weakly.  Thus, we set $U(\Ctilde_g,\frac{\tau}{n})$ to the identity and simplify Eq. (\ref{eq:utrotter}) to
\begin{equation}
U_g\rp{\tau} \approx \prod_kU\rp{\C_g^k,\tau}.
\end{equation}
This approximation requires independent calculation of propagators for each group $g$, which corresponds to $N/g$ independent calculations of $2^g \times 2^g$ matrices, exponentially less
difficult than the original problem of direct calculation of the $2^N \times 2^N$ dimensional matrix.  We remark that including the effect of $U(\Ctilde_g,\frac{\tau}{n})$ in the Trotterization can be done by using a tree tensor network ansatz wave function~\cite{shi2006} where the number of complex coefficients to describe the wave function is $O(N^{log(N)})$ instead of $2^N$.

\section{Electron spin-echo}\label{sec:ese}

Electron spin-echo removes static magnetic shifts caused by a spin bath, allowing to measure the dynamical changes of the bath. Assuming that an initial state, $\ket{\varphi} = (\ket{0}+ \ket{1})/\sqrt{2}$, is prepared, the probability of recovering the same state after a time $2\tau$ is 
\begin{equation}
\label{eq:p} p = \mathrm{Tr}(P_{\varphi}U_T(\tau)\rho U_T^{\dag}(\tau)) , 
\end{equation}
where $P_{\varphi} = \ketbra{\varphi}{\varphi}$ is the projector operator to the initial state,
$\rho=\ketbra{\varphi}{\varphi}\otimes\rho_n$ is the density matrix of the total system, $\rho_n$ is the density matrix of the spin bath, $U_T\rp{\tau} = U\rp{\tau}R_{\pi}U\rp{\tau}$ is the total evolution of the system where $U$ is the evolution operator under hamiltonian (\ref{eq:ham}) and $R_\pi$ is a $\pi$-pulse acting on the subspace $m_s=\{0,1\}$ of the electron spin manifold. Probability (\ref{eq:p}) can also be written as $p= (1+S\rp{\tau})/2$, where 
\begin{equation}
\label{eq:ps}%
S\rp{\tau} = \mathrm{Tr}\rp{\rho_n
U_0^{\dag}\rp{\tau}U_1^{\dag}\rp{\tau}U_0\rp{\tau}U_1\rp{\tau}}
\end{equation}%
is known as the pseudo spin and $|S(\tau)|=0$ is the long-time (completely decohered) signal. In the high temperature limit, the density matrix of the nuclei can be approximated by $\rho_n\approx\mathbbm{1}^{\otimes N}/2^N$ where $N$ is the number of nuclei. The generalization of this relation for different sublevels of the triplet state is straighforward, $S\rp{\tau} = \mathrm{Tr}\rp{\rho_n U_\alpha^{\dag}\rp{\tau}U_\beta^{\dag}\rp{\tau}U_\alpha\rp{\tau}U_\beta\rp{\tau}}$, where $\alpha=1$ and $\beta=-1$, for example. In what follows, we analyze the effect of an interacting bath on Eqn. (\ref{eq:ps}).

\subsection{Non-interacting bath}\label{sec:singlebody}

To understand the effect of an interacting bath we will first analyze the non-interacting case, which displays the phenomenon of electron spin-echo envelope modulation due to electron spin-nuclear spin entanglement. This is completely neglecting interactions among nuclei, $\mathbf{C}_{nm}=0$. In this regime, the evolution operator is factored out for each nucleus and the pseudo-spin is the product of all single pseudo-spin relations. In the high temperature limit, $\rho_n=\mathbbm{1}/2$, we obtain the exact expression\cite{childress2006}
\begin{eqnarray} \label{eq:nips10}
S_T\rp{\tau} &=& \prod_n S_n\rp{\tau}= \prod_n
\rp{1-2\bp{\hat{\Omega}_{n}^{(0)}\times\hat{\Omega}_{n}^{(1)}}^2 \right. \nonumber \\
&&\times\left.\sin^2\frac{\Omega_{n}^{(0)}\tau}{2}\sin^2\frac{\Omega_{n}^{(1)}\tau}{2}}.
\end{eqnarray}%

When the electron spin is in its $m_s=0$ state and the external magnetic field points parallel to the NV-axis, the Larmor frequency $\Omega_{n}^{(0)}$ is set by the external magnetic field, and the nuclei precess with the same frequency $\Omega^{(0)}$. The total pseudo spin is $1$ at times $\Omega^{(0)}\tau = 2m\pi$ with $m$ integer. When the electron is in its $m_s=1$ state, the Larmor vector $\mathbf{\Omega}_{n}^{(1)}$ has a contact and dipolar contribution from the hyperfine interaction
$\mathbf{A}_n$ that may point in different directions depending on the position of the nucleus. As a consequence, when interactions from all nuclei are considered, these electron-nuclear dynamics makes the total pseudo-spin relation collapse and revive. However, it does not show any decay of the revival peaks.

We point out that when the transverse (perpendicular to the NV-axis) magnetic field is non-zero, nuclei near the center experience an enhancement in their g-factors leading to a position-dependent
Larmor frequency $\Omega_n^{(0)}$ (see the Appendix). This will result in an effective decay of the signal since the electron state will not be refocused at the same time for all nuclei.

\subsection{Interacting bath: an example}\label{sec:twobody}

When the intra-bath interactions are considered, the spin-echo signal can show decay in addition to the electron-nuclear dynamics. As an illustrative and simple example, consider a pair of nuclei with their Larmor vectors pointing in the same direction regardless the electron spin state (in this case there
is no electron-nuclear dynamics and the non-interacting pseudo-spin relation for two nuclei is $S_{nm} = S_nS_m=1$ (see Eq. \ref{eq:nips10})).  When the interaction between nuclei is included, the pseudo-spin relation can be worked out exactly,
\begin{eqnarray}
\label{eq:ips10}%
S_{nm}\rp{\tau} &=& 1-\sqp{\frac{ \Delta \Omega_{nm}^{0}c_{nm}^{1}-
\Delta \Omega_{nm}^{1}c_{nm}^{0}}{2}}^2\nonumber\\
&& \times
\frac{\sin^2\rp{\omega_{nm}^0\tau}\sin^2\rp{\omega_{nm}^1\tau}}
{\rp{\omega_{nm}^0}^2\rp{\omega_{nm}^1}^2},
\end{eqnarray}%
where $\rp{\omega_{nm}^{m_s}}^2 = \rp{\Delta \Omega_{nm}^{m_s}/2}^2
+ \rp{c_{nm}^{m_s}}^2$, $\Delta \Omega_{nm}^{m_s} = \Omega_{n}^{m_s}
- \Omega_{m}^{m_s}$ and $c_{nm}^{m_s}$ is the strength of the
dipolar interaction
$c_{nm}^{m_s}(\hat{I}_{n+}\hat{I}_{m-}+\hat{I}_{n-}\hat{I}_{m+}-4\hat{I}_{nz}\hat{I}_{mz})$
between nuclei $n$ and $m$. The two frequencies involved in
(\ref{eq:ips10}), $\omega_{nm}^{0}$ and $\omega_{nm}^{1}$, are not necessarily the same for different  pairs of nuclei. They depend on the relative position between nuclei and the relative position of each nucleus to the NV center. Therefore, when all pair interactions are included the pseudo-spin relation decays. In the following section we present an approach to incorporates not only this two body interaction but also $n$-body interactions with $n\le6$. 

\section{the disjoint cluster approach}

The many-body problem can be readily simplified by following the approximation described in section \ref{sec:mbi}. When the interactions that connects different groups are neglected, the evolution operator is factored out in groups and the spin-echo relation becomes simply
\begin{equation}
\label{eq:psg}%
S_g\rp{\tau} \approx \prod_k S\rp{\C_g^k,\tau},
\end{equation}
where $S\rp{\C_g^k,\tau}$ is the pseudo-spin relation, Eqn. (\ref{eq:ps}), for group $\C_g^k$. $S_g\rp{\tau}$ can be calculated numerically and exactly for small $g$ ($\lesssim10$). Therefore,
electron-nuclear and intra-bath hamiltonians can be simultaneously considered.

In the following section, we present our algorithm for sorting strongly interacting nuclei in a random distributed spin bath into well defined groupings. We take the electron spin-echo signal, with the initial state $\ket{\phi} = (\ket{0} + \ket{1})/\sqrt{2}$ as a figure of merit.  We examine the convergence of our disjoint cluster approach as a function of the maximum group size $g$ and consider the statistics of spin-echo for a variety of physical parameters such as Carbon-13 abundance and magnetic field magnitude and orientation.

\subsection{Grouping algorithm}

One of the criteria to aggregate groups of spins is to consider the strength of the intra-bath interaction. This parameter can be summarized in one variable $\mathbb{C}\rp{i,j}$ which is a scalar function of the interaction $\mathbf{C}^{ij}$ between nuclei $i$ and $j$. The aggregation algorithm used for this criterion is as follow. Consider an array $\mathbb{A}$ containing the criterion for all pairs ordered from high to low values in $\mathbb{C}$ and let $\kp{i,j}_n$ be the $n$-th nuclear pair in array $\mathbb{A}$. The array $\mathbb{A}$ is scanned completely and one of the following cases applies for each pair
$\kp{i,j}_n$ 

\begin{itemize}
  \item {\bf if nuclei $i$ \& $j$ belong to different groups:} join both
  groups if $\mathcal{N}\rp{\mathbb{G}(i)}+\mathcal{N}\rp{\mathbb{G}\rp{j}}\leq
g$.
  \item {\bf if nucleus $i$ belong to group $\mathbb{G}(i)$ and nucleus $j$ does not belong to any
group:} add $j$ to group $\mathbb{G}(i)$ if
$\mathcal{N}\rp{\mathbb{G}(i)}<g$. If not, make a new group with
$j$.
  \item {\bf if nucleus $j$ belong to group $\mathbb{G}(j)$ and nucleus $i$ does not belong to any
group:} add $i$ to group $\mathbb{G}(j)$ if
$\mathcal{N}\rp{\mathbb{G}(j)}<g$. If not, make a new group with
$i$.
  \item {\bf if nuclei $i$ \& $j$ do not belong to any group:} make a new group with $i$ \&
  $j$,
\end{itemize}
where $\mathcal{N}\rp{\mathbb{G}}$ is the number of nuclei in group $\mathbb{G}$ and $g$ is the maximum number of nuclei per group. In what follows, we use the criterion $\mathbb{C}\rp{i,j} =
(C_{xx}^{ij})^2+(C_{yy}^{ij})^2+(C_{zz}^{ij})^2$ (i.e. the interaction between nuclei $i$ and $j$) to estimate the electron spin-echo in NV centers.

\subsection{Numerical methods and example cases}

\begin{figure}[t]
\includegraphics[width=0.95\columnwidth]{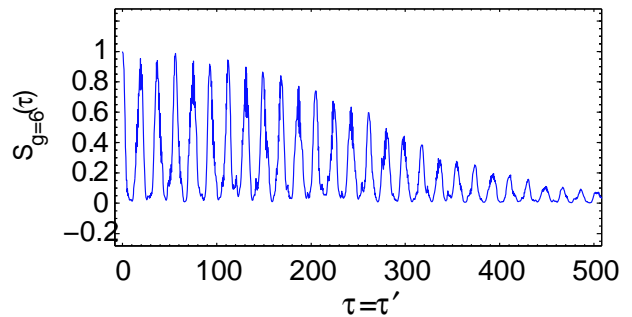}
\caption{ Simulation of the Pseudo spin $S_{g=6}(\tau)$ for a single
NV center in a magnetic field of 50 Gauss oriented parallel the NV-axis.}\label{fig:theo}
\end{figure}

Figure \ref{fig:theo} shows $S_{g}\rp{\tau}$ for $g=6$ (Eq. (\ref{eq:psg})) for 750 random and distributed Carbon-13 in a diamond lattice in a magnetic field of 50 Gauss oriented along the NV-axis. The algorithm was implemented using MATLAB and the Hamiltonian for each group was diagonalized exactly followed by the calculation of the corresponding unitary matrices for 6000 points from 0 to 1 ms.
Each simulation of $S_{g}\rp{\tau}$ takes approximately 10 min. 

\begin{figure}[t]
\includegraphics[width=1\columnwidth]{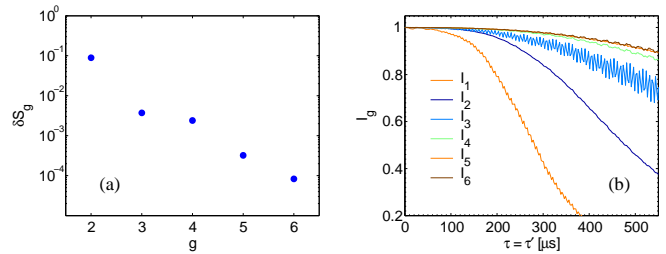}
\caption{(a) Convergence: equation (\ref{eq:dg}) as the maximum number of nuclei $g$ per group is increased. (b) Indicators $I_g$ of the contribution of neglected pairs. When $g$ is increased, the most
important pair interactions are added to the pseudo spin relation $S_g$. The rest is used to calculate $I_g$.} \label{fig:conv} 
\end{figure}

The method also shows good convergence. When the maximum size of subgroups $g$ is increased, more interactions among nuclei are considered and the approximation gets better. As a figure of merit,
we plot the integrated squared difference between consecutive spin-echo relations, $S_g$ and $S_{g-1}$,
\begin{equation}\label{eq:dg}
\pro{\delta S_g^2}  = \frac{1}{T}\int_0^T \sqp{S_g\rp{t} -
S_{g-1}\rp{t}}^2dt.
\end{equation}
Figure \ref{fig:conv}a shows $\frac{1}{2}\log{\pro{\delta S_g^2}}$ up to $g=6$. Each time the maximum size $g$ of subgroups $\C_g^k$ is increased, the spin echo relations, $S_g$'s, get closer.

In addition, following Ref.~\onlinecite{witzel2005}, we introduce the following indicator of all interactions not included in groups $\C_g^k$, and therefore in $S_g$,
\begin{equation}
\label{eq:ind}%
I_g\rp{\tau} = \prod_{\kp{n,m}\epsilon\Ctilde_g}S_{nm}\rp{\tau}.
\end{equation}
The product in Eq. (\ref{eq:ind}) runs over all neglected pair interactions contained in $\Ctilde_g$ and $S_{nm}$ is calculated according to Eq. (\ref{eq:ips10}). $I_g\rp{\tau}$, which obeys $0\leq I_g\rp{\tau}\leq 1$, is an indicator of convergence for our approach. When $I_g\rp{\tau}$ is close to unity, good convergence is achieved. Figure \ref{fig:conv}b shows $I_g\rp{\tau}$ for several aggregations (different $g$'s). As expected, when the maximum subgroup size $g$ is increased, the contribution from all neglected interactions is small. By the time the neglected interactions become important, the pseudo-spin $S_g\rp{\tau}$ has already decayed (see figure \ref{fig:theo}).

\section{Results and Discussion}

\begin{figure}[t]
\includegraphics[width=1\columnwidth]{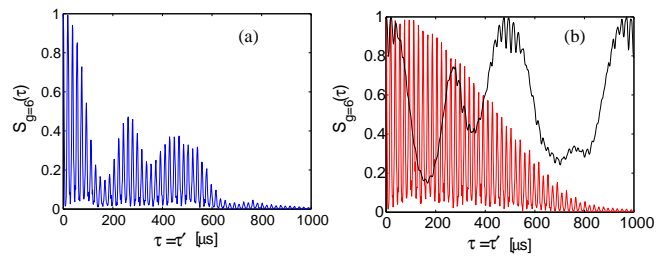}
\caption{(a) Electron spin-echo signal highly modulated by a few Carbon-13 that strongly interact with the electron spin. (b) The strong contribution to the signal (black curve) has been isolated from the contribution from the rest of the spin bath (red curve).}\label{fig:mod}
\end{figure}

The results shown in figure \ref{fig:mod} clearly indicates that the electron spin echo signal cannot be modeled by just one time scale. This result can be understood by noting that few strongly interacting nuclei can coherently modulate the usual exponential decay. This is in good quantitative agreement with recent experimental results\cite{maze2008nature}.

The random distribution of the spin bath and the relative high coupling between two nearest neighbor nuclei ($\sim2$ kHz) may cause a few nuclei to contribute significantly to the decay of the spin-echo signal. Nuclei that makes small contributions to the decoherence of the electron contribute as $1-a\tau^4\approx\exp(-a\tau^4)$ as it can be seen from Eq. (\ref{eq:ips10}). This behavior starts to deviate from $\exp(-a\tau^4)$ as the interaction between nuclei increases. Figure \ref{fig:mod}a shows a very unusual decay at which few nuclei modulate coherently (Fig. \ref{fig:mod}b, black curve) the irreversible contribution from the rest of the bath (Fig. \ref{fig:mod}b, red curve). Therefore, individual NV centers can show a rich variety of spin-echo signals with multiple time scales. The coherent modulation of the spin-echo diffusion due to strong interacting nuclei suggests that we can think about a system composed of the electron and these few strong interacting nuclei and an environment composed of the rest of the spin-bath.

\begin{figure}[t]
\includegraphics[width=1\columnwidth]{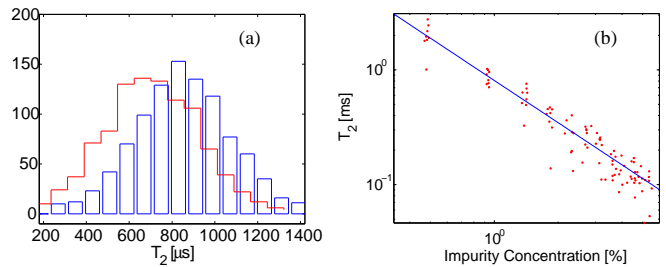}
\caption{Histogram of $T_2$ for 1000 simulations at a magnetic field of $50$ Gauss at an angle of
$\theta= 0^\circ$ (blue) and $\theta= 6^\circ$ (red) with respect to the NV-axis. (b) Decoherence time $T_2$ versus impurity concentration, Carbon-13, at 50 Gauss along the NV axis.}\label{fig:hist}
\end{figure}

Each NV center experiences a different random configuration and concentration of Carbon-13. This causes a large distribution of decoherence times $T_2$ when many centers are probed. In order to estimate the decoherence time $T_2$ we fit the envelope of $S_g\rp{\tau}$ to $\exp\rp{-(2\tau/T_2)^3}$. When the fit is not accurate we define $T_2$ as the longest time for which $S_g\geq1/e$. Figure
\ref{fig:hist}a shows the histogram of $T_2$ for 1000 different random distribution of Carbon-13 in the diamond lattices for an external magnetic field of 50 Gauss
parallel to the NV-axis.

For non-zero transverse magnetic fields, the electron-nuclear interaction, Eq. (\ref{eq:nips10}), makes an important contribution due to the enhancement in the nuclear g-factor and interaction between nuclei (see Eq. (\ref{eq:ccc})). In this case, nuclei will have different Larmor frequencies $\Omega^0_{n}$ depending on the relative distance between them and the center. Each nuclei will refocus at different times, decreasing the maxima of the revivals, i.e., decay of the signal. This is illustrated in figure \ref{fig:hist}a for a magnetic field at an angle of $\theta=6^\circ$ from the NV-axis. 

As expected, the decoherence time decreases as the impurity concentration increases. This is shown in figure \ref{fig:hist}b where $T_2$ goes as $1/n$. To understand this it is possible to make an analysis using a small $\tau$ expansion; while this is not always correct, it provides a simple explanation of the underlying behavior. From Eq. (\ref{eq:ips10}), the decoherence time scales as the geometric mean of the bath dynamics and the bath-spin interaction, i.e., $T_2\sim\rp{\bar{C}A_c}^{-1/2}$, where $\bar{C}$ is the averaged nuclear-nuclear dipolar interaction and $A_c$ is some characteristic value for the electron-nuclear interaction. Since both interactions decay as $r^{-3}$ and the average distance between bodies scales with the concentration as $n^{-1/3}$, both interactions scale linearly in $n$. Therefore, the decoherence time $T_2$ decreases approximately as $1/n$.

\begin{figure}[t]
\includegraphics[width=1\columnwidth]{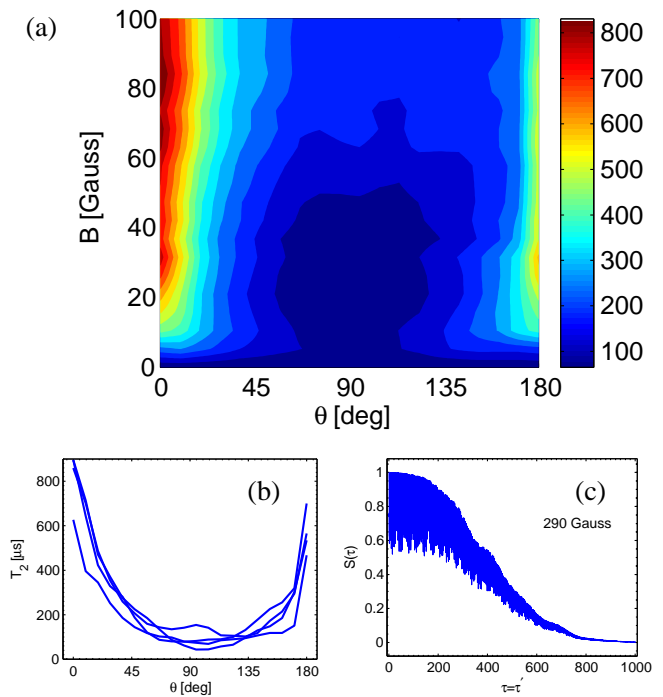}
\caption{(a) Coherent time $T_2$ for different magnetic field strength and angles (measured from the NV-axis). Each point is average over 6 different random distributed baths. (b) Coherence time $T_2$ versus angle of the magnetic field for 4 different spin baths at 50 Gauss. (c) Pseudo spin $S_{g=6}\rp{\tau}$ at 290 Gauss. At high fields the collapses due to the electron-nuclear dynamics decreases (see text).}\label{fig:map}
\end{figure}

As the angle between the magnetic field and the NV-axis, $\theta$, is increased, the electron-nuclear dynamics dominates and the spin-echo signal shows small revivals and fits poorly to a single exponential decay. Thus, to describe the coherence time at these angles, we have plot the average value of the signal, normalized by the average signal at $\theta=0^{\circ}$:\footnote{
    We choose this approach as the signal has sufficiently non-trivial
    electron-nuclear dynamics for $\theta \neq 0$ that fitting an envelope
    decay function, as is done for $\theta=0$, leads to large errors.}
\begin{equation}
T_2(B,\theta) \equiv T_2(B,\theta=0) \frac{\int_0^\infty |S_{B,\theta}(t)| dt}{\int_0^\infty
  |S_{B,\theta=0}(t)| dt}\ .
\end{equation}
Figure \ref{fig:map}a shows how the coherence of the signal varies with the strength and orientation of the magnetic field. This map is averaged over 6 different spin baths, since the random localization of Carbon-13 nuclei in the lattice makes the coherent time to vary from NV-center to NV-center as it can be seen in figure \ref{fig:map}b for a fixed magnetic field.

When the magnetic field along the NV-axis increases, the contribution from the electron-nuclear interactions decreases (see figure \ref{fig:map}c). This happens because the quantization axis for the
nuclei points almost in the direction of the external magnetic field producing a small oscillating field. This can be easily seen in the non-interacting case, Eq. (\ref{eq:nips10}), where the second term vanishes if $\mathbf{\Omega}^{(0)}_n \parallel \mathbf{\Omega}^{(1)}_n$. Similarly, when electron spin-echo is performed using the sub manifold $m_s=\{+1,-1\}$, the signal does not revive since each nuclei refocus the electron at different times. This occurs because the Larmor frequencies in this case, $\Omega^{\pm1}_n$, are position dependent and differ for each nuclei.

We also point out that the approximation introduced in section \ref{sec:mbi} is valid as long as the impurity concentration of Carbon-13 is not too high, so the neglected interactions that connect different groups do not play an important role. This allows us to treat the bath as isolated groups. The approximation also relies on the relatively large interaction between the central spin (electron) and the bath, $\mathbf{A}_n$, when compared to the intra-bath interaction, $\mathbf{C}_{nm}$. The reason for this is that as the central spin gets disconnected from the bath (reducing $\mathbf{A}_n$), the decay occurs at later times $\tau$ and interactions of the order of $\tau^{-1}$ start to play a role. To
illustrate this, consider the size of each subgroup scaling as $\rp{g/n}^{1/3}$ where $g$ is the size of the subgroup. Then, the interaction between nearest neighbor groups scales as $nC_{nn}/g$
where $C_{nn}$ is the nearest neighbor nuclear interaction. The time at which this interaction is important goes as $t\sim g/nC_{nn}=g/\bar{C}$. If we require this time to be larger than the decoherence time ($t\gg T_2$), we find that the two types of interactions should satisfy $g\rp{A_c/\bar{C}}^{1/2}\gg1$ (for the present study this value is around 150). Therefore, when the interaction between the addressed
spin and the bath is of the order of the intra-bath interactions, the approximation breaks down. This would be the case of the spin-echo signal for a nuclear spin proximal to the NV center\cite{dutt2007} in which more sophisticated methods should be applied such as tree tensor networks\cite{shi2006}.

\section{Conclusions}

We have presented a method to evaluate the decoherence of a single spin in the presence of an interacting randomly-distributed bath. It properly incorporates the strong electron-nuclear dynamics present in NV centers and explains how it affects the decoherence. We also incorporates the dynamic beyond the secular approximation by including an enhanced nuclear g-factor that depend on the orientation of the external magnetic field relative to the NV-axis and by including an electron mediated nuclei interaction. Our results show that the spin echo signal for NV centers can present multiple time scales where the exponential decay produced by many small nuclei contributions can be coherently modulated by few strongly interacting nuclei. The coherence times in ultra-pure diamond can be further improved by making isotopically pure diamond with low concentration in Carbon-13. This method may be used in other systems as long as the intra-bath interaction is smaller than the interaction between the central spin and the bath. These results have important implications, e.g., in magnetometry where long coherence times are important. For example, echo signals persisting for up to milliseconds can be used for nanoscale sensing of weak magnetic fields, as it was demonstrated recently\cite{maze2008nature}.

\appendix
\section{Interactions for NV centers}\label{app:int}

Electron spin resonance shows that the nuclear spin-electron spin
interaction is $150$ MHz\cite{he1993} for the three nearest neighbor
Carbon-13's and around 2 MHz for the Nitrogen. Away from this deep
defect, the interaction is dipolar-like,
\begin{equation}
\mathbf{S} \cdot \mathbf{A}_n \cdot \mathbf{I}_n \equiv 5.6\rp{\frac{a_{nn}}{R_n}}^3\rp{3(\hat{S}\cdot
  \vec{n})(\hat{I}_n\cdot \vec{n}) - \hat{S}\cdot \hat{I}_n} {\rm
  MHz}
\end{equation}
where $a_{nn} = 1.54{\AA}$ is the nearest neighbor distance for
diamond, $R_n$ is the distance between the $n$-th Carbon-13 and the
defect, and $\vec{n}$ is the unit vector that connects the electron
and the nucleus. Carbon-13's interact via dipolar interaction,
\begin{equation}
\mathbf{I}_n \cdot \mathbf{C}_{nm} \cdot \mathbf{I}_m \equiv 2.1\rp{\frac{a_{nn}}{R_{nm}}}^3\rp{3(\hat{I}_n\cdot
  \vec{n})(\hat{I}_m\cdot \vec{n}) - \hat{I}_n\cdot \hat{I}_m} {\rm
  kHz,}
\end{equation}
where $R_{nm}$ is the distance between the $n$-th and $m$-th
Carbon-13.

Recent experiments have been performed at fields of $1-100$
Gauss\cite{childress2006} with Zeeman energies of $\sim$MHz and
$\sim$kHz for the electron and Carbon-13 nuclear spins, respectively. At
these fields, it is convenient to make the so-called secular
approximation thanks to the large splitting $\Delta$. However,
transverse directions (perpendicular to the NV-axis) of the external
magnetic field and hyperfine fields from Carbon-13 must be taken
into account as a perturbation to get an accurate description of the
system, as it has been shown by experiments\cite{childress2006}.
Including this, we can write the hamiltonian in the following form
\begin{eqnarray}
\label{eq:sham}%
H & \approx & \Delta S_z^2 - \gamma_eB_zS_z +
\sum_{nj}S_zA_{zj}^nI_j^n +\sum_n
\delta\mathbf{A}_n^{\mathrm{T}}\cdot\mathbf{I}_n \nonumber\\
&&-\gamma_N\sum_n\textbf{B}^{\mathrm{T}}\cdot\mathbf{g}^{\mathrm{eff}}_n\cdot\textbf{I}_n
+\sum_{n>m}\mathbf{I}_n^{\mathrm{T}}\cdot\mathbf{C}^{\mathrm{eff}}_{nm}\cdot\mathbf{I}_m
\end{eqnarray}
where $\mathbf{g}^{\mathrm{eff}}_n  =  1 + \delta
\mathbf{g}_n\rp{m_s}$ is the effective g-tensor\cite{childress2006},
$\mathbf{C}^{\mathrm{eff}}_{nm}  =  \mathbf{C}_{nm} +
\delta\mathbf{C}_{nm}\rp{m_s}$ is the effective coupling between
Carbon-13's and
\begin{eqnarray}
\label{eq:ccc}%
\delta \mathbf{g}_n\rp{m_s} & = & -
\frac{\rp{2-3|m_s|}\gamma_e}{\Delta\gamma_N}\rp{\begin{array}{ccc}
                                         A_{xx}^n & A_{xy}^n & A_{xz}^n \\
                                         A_{yx}^n & A_{yy}^n & A_{yz}^n\\
                                         0 & 0 & 0
                                       \end{array}
}\nonumber\\
\delta \mathbf{C}_{nm}\rp{m_s} & = & -
\frac{\rp{\Delta\gamma_N/\gamma_e}^2}{\Delta\rp{2-3|m_s|}}
\delta{\mathbf{g}_n^{T}}\cdot\delta{\mathbf{g}_m} \\
\delta\mathbf{A}_n^{\mathrm{T}}\rp{m_s} & = &
\frac{\rp{2-3|m_s|}}{2\Delta}\sum_{ij}\epsilon_{ijk}A_{xi}^nA_{yj}^n.
\nonumber
\end{eqnarray}
For nuclei close to the center, $\delta g$ can reach values between
$0-15$\cite{childress2006}, $\delta C$ can be several times the bare
dipole-dipole interaction\cite{dutt2007}. The term $\delta A$ is
a small contribution that can be neglected in most of the cases due
to the large value of the zero-field splitting $\Delta$.

\begin{acknowledgments}
The authors would like to thank S. Das Sarma for useful comments. This work was supported by NSF, DARPA and Packard foundation. JRM thanks Fulbright-Conicyt scholarship for support. JMT is supported by a Pappalardo Fellowship.
\end{acknowledgments}


\end{document}